\documentclass[sigconf]{acmart}
\AtBeginDocument{%
  \providecommand\BibTeX{{%
    \normalfont B\kern-0.5em{\scshape i\kern-0.25em b}\kern-0.8em\TeX}}}

\setcopyright{acmcopyright}
\copyrightyear{2023}
\acmYear{2023}
\acmDOI{10.1145/3581641.3584078}

\acmConference[IUI '23]{28th International Conference on Intelligent User Interfaces}{March 27--31, 2023}{Sydney, NSW, Australia}
\acmBooktitle{28th International Conference on Intelligent User Interfaces (IUI '23), March 27--31, 2023, Sydney, NSW, Australia}
\acmPrice{15.00}
\acmISBN{979-8-4007-0106-1/23/03}

\usepackage{multicol}
\usepackage{multirow}
\usepackage{adjustbox}
\usepackage{xspace}
\usepackage{xcolor}
\usepackage{seqsplit}

\begin{document}

\title{Large-scale Text-to-Image Generation Models for Visual Artists' Creative Works}

\author{Hyung-Kwon Ko}
\affiliation{%
  \institution{KAIST}
  \country{Republic of Korea}
}
\email{hyungkwonko@gmail.com}

\author{Gwanmo Park}
\affiliation{%
  \institution{Seoul National University}
  \country{Republic of Korea}
}
\email{parkgw95@gmail.com}

\author{Hyeon Jeon}
\affiliation{%
  \institution{Seoul National University}
  \country{Republic of Korea}
}
\email{hj@hcil.snu.ac.kr}

\author{Jaemin Jo}
\affiliation{%
  \institution{Sungkyunkwan University}
  \country{Republic of Korea}
}
\email{jmjo@skku.edu}

\author{Juho Kim}
\affiliation{%
  \institution{KAIST}
  \country{Republic of Korea}
}
\email{juhokim@kaist.ac.kr}

\author{Jinwook Seo}
\authornote{Corresponding author}
\affiliation{%
  \institution{Seoul National University}
  \country{Republic of Korea}
}
\email{jseo@snu.ac.kr}






\renewcommand{\shortauthors}{Ko et al.}

\begin{abstract}
Large-scale Text-to-image Generation Models (LTGMs) (e.g., DALL-E), self-supervised deep learning models trained on a huge dataset, have demonstrated the capacity for generating high-quality open-domain images from multi-modal input.
Although they can even produce anthropomorphized versions of objects and animals, combine irrelevant concepts in reasonable ways, and give variation to any user-provided images, we witnessed such rapid technological advancement left many visual artists disoriented in leveraging LTGMs more actively in their creative works.
Our goal in this work is to understand how visual artists would adopt LTGMs to support their creative works.
To this end, we conducted an interview study as well as a systematic literature review of 72 system/application papers for a thorough examination.
A total of 28 visual artists covering 35 distinct visual art domains acknowledged LTGMs' versatile roles with high usability to support creative works in automating the creation process (i.e., automation), expanding their ideas (i.e., exploration), and facilitating or arbitrating in communication (i.e., mediation).
We conclude by providing four design guidelines that future researchers can refer to in making intelligent user interfaces using LTGMs.

\end{abstract}



\renewcommand{\subsectionautorefname}{Section}
\renewcommand{\sectionautorefname}{Section}
\newcommand{\grayrow}{\rowcolor{gray!13}}

\maketitle

\section{Introduction}

Large-scale Text-to-image Generation Models (LTGM) (e.g., DALL-E~\cite{ramesh2021zero, ramesh2022hierarchical}) are AI models trained at scale (e.g., 250 million text-images pairs for DALL-E) that show generalizable performance at several downstream tasks such as image captioning and object detection \cite{bommasani2021opportunities}.
Compared to previous AI models, the biggest advantage of LTGMs is that they can take multi-modal input, such as text or image, to produce high-quality images in a zero-shot fashion \cite{ramesh2021zero}.
For example, DALL-E even achieved high generation quality when tested on MS-COCO dataset~\cite{lin2014microsoft}, although it was not contained in the training labels \cite{ramesh2021zero}.
Thanks to the advancement of LTGMs, there exists multiple software to make artworks \cite{wombo, midjourney, artbreeder} called AI art generator, which has drawn huge attention from various visual art domains \cite{maeda, lieberman, klingemann}.


In the visual art domain, AI art generators have gained popularity and found versatile usage.
More and more visual artists expressed their interests toward them on social media platforms.
For example, John Maeda, a former president at Rhode Island School of Design, said LTGMs can become a tool to change the working paradigm of visual artists by taking initiatives in producing art, even performing better than humans \cite{maeda}.
However, we observed that rapid technological improvements have left many visual artists perplexed and disoriented to ponder on how to adopt LTGM actively to their creative works.
Moreover, while prior work has highlighted LTGMs' technical properties such as prompt engineering \cite{liu2022design, 10.1145/3527927.3532792}, yet little research has explored the applicability of LTGMs in assisting visual artists broadly.
In this work, we seek to answer the following research questions: 1) which subgroups of visual artists would be more willing to use LTGMs? 2) what types of tasks would they utilize LTGMs on? 3) what would the LTGM's role be in those cases?

\begin{figure*}[t]
  \centering
  \includegraphics[width=\linewidth]{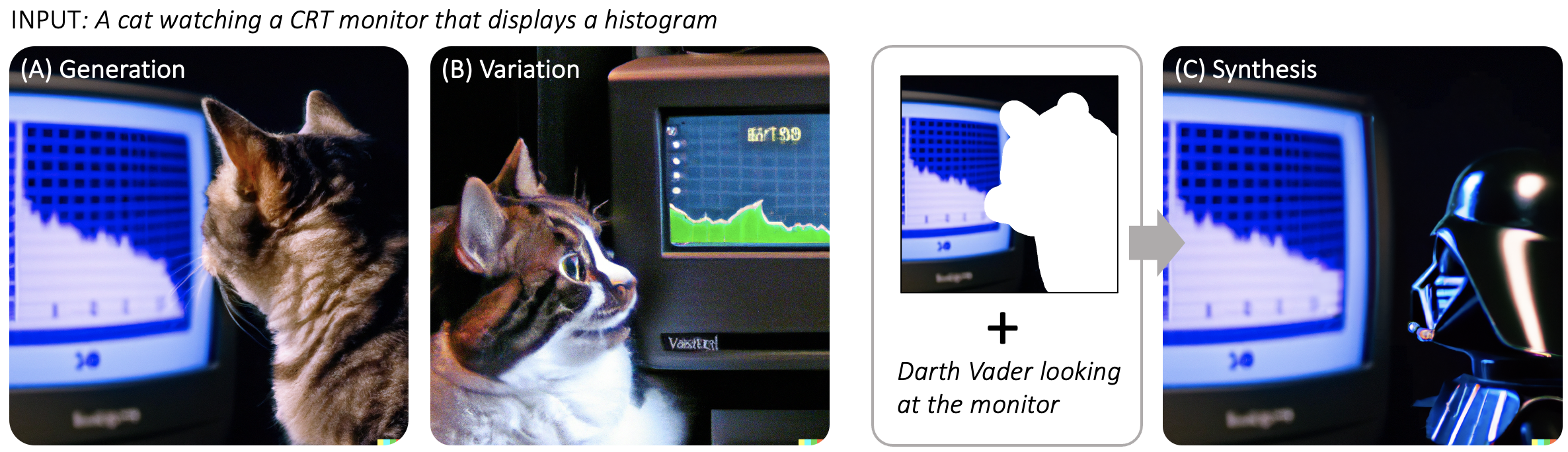}
  \caption{\textbf{Example images of LTGM with DALL-E's three functionalities.} (A) A generated image by DALL-E from a text prompt \textit{"A cat watching a CRT monitor that displays a histogram"}. (B) A variation of the generated image. (C) Providing both an image with brushed region and text can synthesize a new image that matches with given inputs. In this case, we erased the cat and changed it to Darth Vader.}
  \label{fig:dalle}
\end{figure*}


We conducted a systematic literature review of 72 system and application papers as well as a semi-structured interview with 28 visual artists in varying domains (e.g., fine art and applied art) to find answers to the research questions.
The interview study was grounded in three meaningful themes (i.e., user, task, and role) found in the literature review, and they were used to examine the findings afterward.
The study results showed that LTGMs can perform diverse roles including 1) automating the creation process (i.e., automation), 2) expanding their ideas (i.e., exploration), and 3) facilitating or arbitrating in communication (i.e., mediation).
However, we also found that visual artists found it hard to actually incorporate LTGMs into their creative works in its current form.




Based on the results of our interview study, we suggest four design guidelines for researchers and practitioners interested in making intelligent user interfaces using LTGMs.
To elicit LTGMs' potential usability to the maximum, we propose to 1) support variability level specification for different types of visual art, 2) provide model customization grounded in domain-specific understanding, 3) add more controllability using multi-modal input, and 4) help prompt engineering for the ease of writing text input.

The main contributions of our work are summarized as follows:
\begin{itemize}
  \item We conducted a systematic literature review of 72 system and application papers that used generative models to understand the context (e.g., user, task, and role) of how they have been used in the HCI domain;
  \item We conducted an interview study of 28 visual artists with varying occupations to investigate how visual artists would adopt LTGMs in their domain to support their creative works;
  \item We provide design guidelines that future researchers can refer to in building intelligent user interfaces using LTGM.
\end{itemize}







\begin{figure}[t]
  \centering
  \includegraphics[width=\linewidth]{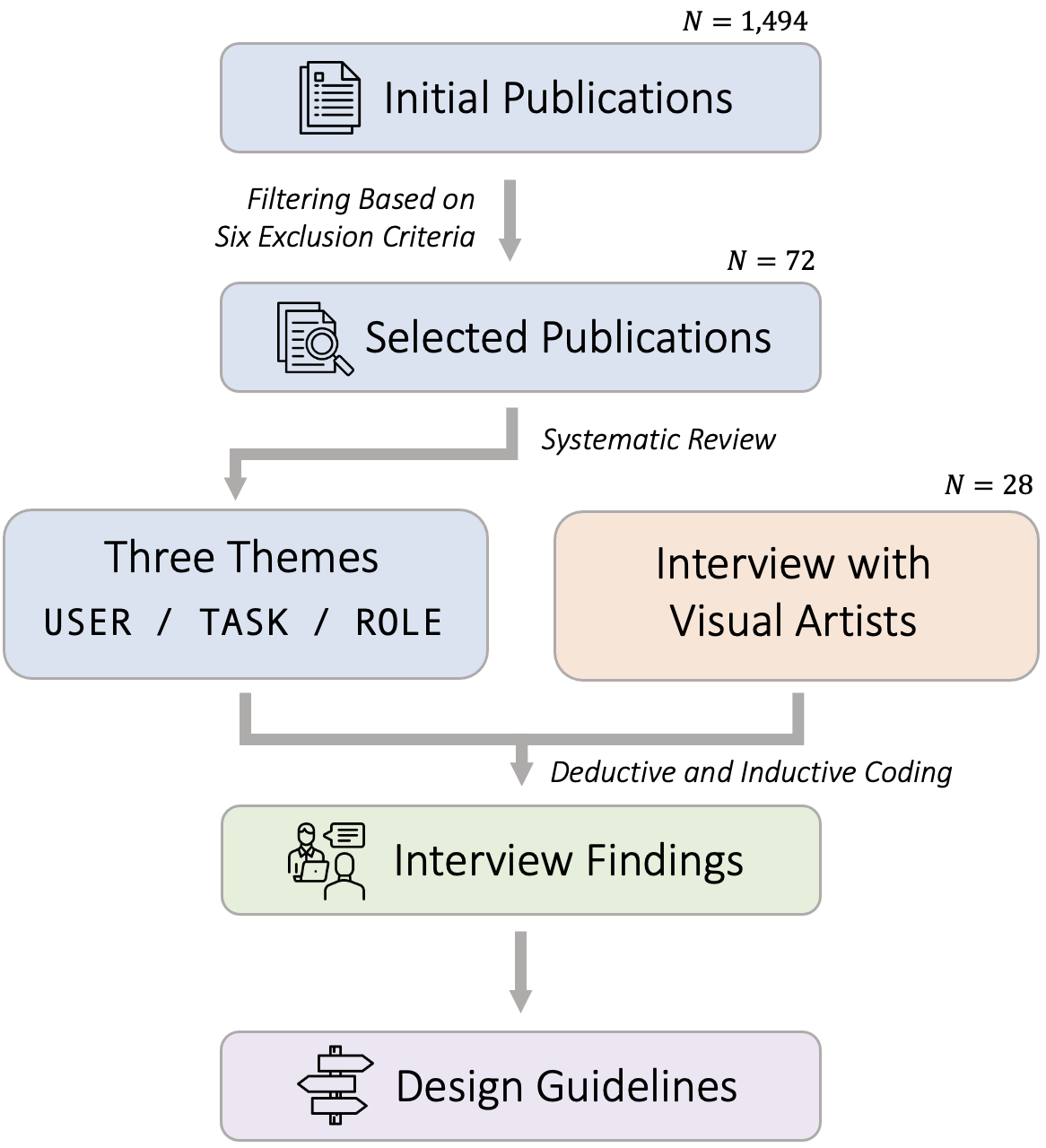}
  \caption{\textbf{The overall study and analysis procedure.} First, we build themes and codes based on a systematic literature review of 72 publications (\autoref{sec3}). Then we analyze the data of interviews with 28 visual artists based on both the deductive and inductive coding process (\autoref{sec4} and \autoref{sec5}). Finally, we propose four design guidelines for future researchers (\autoref{sec6}).}
  \label{fig:overview}
\end{figure}

\section{Background and Related Work}

In this section, we first introduce the large-scale text-to-image generation models (LTGMs) and their democratization.
Then, we present how AI models have been understood as a tool to support visual artists' creative works in the HCI domain.


\subsection{LTGMs and the Impact of Their Democratization}

In 2021, OpenAI presented DALL-E~\cite{ramesh2021zero}, and since then more than 109k people joined Reddit community\footnote{https://www.reddit.com/r/dalle2} to post, share and have a discussion on DALL-E-generated images.
DALL-E consists of 12 billion parameters, which is approximately 632 times larger than the preceding text-to-image generation model~\cite{tao2022df}.
While previous models could only generate images dependent on training datasets such as MS-COCO~\cite{lin2014microsoft} or CUB~\cite{wah2011caltech}, DALL-E has shown a generalizable image generation performance even on unseen datasets in terms of both quality metrics and human evaluators.
Moreover, it also offers some other useful functions such as image variation and image synthesis (\autoref{fig:dalle}).
A year after, OpenAI introduced a highly improved model called DALL-E2~\cite{ramesh2022hierarchical} which can generate realistic images and artworks with four times higher resolution (1024 * 1024 pixels) than the previous one (256 * 256 pixels).
The performance of the DALL-E series was so powerful that it affected many other companies to accelerate the development of LTGMs.
Some of them are N{\"U}WA series~\cite{wu2021n, wu2022nuwa}, made by Microsoft, Parti~\cite{yu2022scaling} developed by Google, and Make-A-Scene~\cite{gafni2022make} released by Meta.
In most cases, the model weights and training source codes were not open to the public, which makes it hard for the researchers to reproduce the image generation and editing process on their own.

However, LTGMs become more accessible to the public as many people have put a lot of effort into democratizing it.
Some LTGMs, such as Latent Diffusion Models~\cite{rombach2021highresolution} was trained on open source datasets~\cite{schuhmann2021laion, schuhmann2022laion} and the authors uploaded it on the web so general public or researchers can access them freely, e.g., for further research \cite{gal2022textual}.
Based on such advancement that datasets, source codes and model weights become easily accessible, a huge number of web applications using LTGMs are now flooded to the public.
For example, since LTGMs accept text prompt, there is a prompt marketplace~\cite{promptbase} to buy and sell a pertinent text prompt.
Also, a search engine~\cite{lexica} exists to find more than 5 million LTGM-generated images, which is similar to previous reference search tools like Pinterest\footnote{https://www.pinterest.com}.
Recently, NovelAI presented an anime character generation model with simple brushing interaction, which can ease the burden of character designers \cite{novelai}.
In light of this concentrated attention toward LTGMs, it is reasonable to say that LTGMs will have a massive impact to visual artists who deal with visual source as main component of their work.

\subsection{AI models to Support Visual Artists' Creative Works}
So far, AI has been a useful tool to help visual artists' work \cite{shneiderman2022human}.
There are a lot of AI applications and systems to help experts in different visual art domains such as graphic design \cite{ueno2021continuous}, UI design \cite{wang2021screen2words}, webtoon \cite{wetoon}, digital art \cite{yurman2022drawing}, new media art \cite{10.1145/3527927.3532792}, and so on.
Some of their goals were to enhance the expressivity of 3D faces' facial expressions \cite{abdrashitov2020interactive}, or to automate generating wireframes of chosen UI design patterns \cite{gajjar2021akin}. 
Considering a guideline for creation tools that the researchers should keep a proper moderation between human control and computer automation \cite{shneiderman2020human}, some actively adopted a human-in-the-loop approach to facilitate human-AI collaboration.
In detail, they tried to support augmenting creativity by providing an embodied experience in VR environment \cite{urban2021designing}, and a controllable interface to balance between exploration and exploitation \cite{10.1145/3397481.3450663}.
In most cases, the collaboration was recognized as a positive proxy by helping design ideation \cite{karimi2020creative} and generating variety of sets of outputs efficiently \cite{liu2022opal}.

On the other hand, there are many papers studied artist's overall circumstance or several implications these AI models can bring up.
Chung et al. researched an artist's support network to interpret a complex relationship between artists needed for an art-making \cite{10.1145/3532106.3533505}.
Through this analysis, they tried to inform the design of creativity support tools in connection with the network.
Similarly, Li et al. explored the software’s role in visual art production to inform end-user programming and creativity support tools \cite{li2021we}.
Also, HCI researchers have discussed the challenges and opportunities that AI models have brought as they become a part of creation tools \cite{bommasani2021opportunities}.
Examples are challenges concerning ethical issues such as abuse and legality.
For example, Deepfake has exacerbated a social problem of generating deception, propaganda, and disinformation \cite{tahir2021seeing, gamage2022deepfakes}, which is not even decided in our society that who has the legal responsibility for such fake media \cite{ali2021exploring}.


Although the aforementioned research has discussed specific applications and their diverse implications to support creative works, little has been done regarding the LTGMs' potential to change the way future visual artists work \cite{bommasani2021opportunities}.
Only a few branches of papers has delved into suggesting technical guidelines on how to write a proper text prompt \cite{liu2022design, 10.1145/3527927.3532792}, and how to evaluate the LTGMs so that the researcher can better understand their reasoning process and social biases \cite{cho2022dall}.
Rather than focusing on the technical characteristics of the LTGMs, we concentrate on comprehending how do the people in the visual art domain leverage the LTGMs to help their creative works.
To this end, our goal is to examine what impact it can bring to the way future visual artists work.

\section{Systematic Literature Review}
\label{sec3}

\begin{table}[t]
\caption{\textbf{A summary of the reviewed papers.} After multiple iterations of scrutinizing, 72 papers were chosen for the final analysis.}
\label{tab:review-number}
\begin{center}
\begin{tabular}{lrrrr} 
\toprule
 & Round 1 & Round 2 & Round 3 & \textbf{Round 4} \\ 
\midrule
CHI & 778 & 728 & 25 & 26 \\ 
UIST & 182 & 178 & 4 & 7 \\ 
IUI & 267 & 256 & 18 & 19 \\ 
DIS & 147 & 147 & 4 & 4 \\ 
CSCW & 48 & 48 & 1 & 2 \\ 
C\&C & 72 & 72 & 12 & 13 \\ 
Etc. (e.g., CGF) & 0 & 0 & 0 & 1 \\ 
\midrule
SUM & 1,494 & 1,429 & 64 & \textbf{72} \\ 
\bottomrule
\end{tabular}
\end{center}
\end{table}


\begin{table*}[t]
\caption{\textbf{A summary of themes and codes extracted from the systematic literature review.} Examples or definition, counts (\textit{N}), and percentages (\%) of codes in each theme extracted from an iterative analysis of 72 papers.
`User' is the target group of the research.
`Task' is a type of supported work discussed in each paper.
`Role' explains the reasons for adopting the generative model in each paper.
In case of `User' and `Task', we assigned a single code to one paper.
However, we allocated multiple codes of `Role' to a single paper in some cases, since it was often vague to divide them into a single category.}
\label{tab:review-utr}
\begin{center}
\resizebox{\linewidth}{!}{%
\begin{tabular}{cllcc}
\toprule
    Themes & Codes & Examples/Definition & \textit{N} & \%\\
\midrule
    \multirow{9}{*}{User} & \textbf{Artists} & \textbf{Digital artists \cite{yurman2022drawing}; Computational artists \cite{wang2020latent}; 3D modelers \cite{abdrashitov2020interactive}; Webtoon authors \cite{wetoon}}  & \textbf{5} & \textbf{6.9}\\
    & Children/Students & 6- to 10-year old children \cite{zhang2022storydrawer}; Undergraduate students \cite{jonsson2022cracking} & 3 & 4.2\\
    & \textbf{Designers} & \textbf{UI/UX designers \cite{gajjar2021akin, ang2021learning, liu2018learning, wang2021screen2words, mozaffari2022ganspiration}; GUI designers \cite{li2021screen2vec}; Fashion designers \cite{padiyath2021desainer}} & \textbf{13} & \textbf{18.1}\\
    & Engineers & Software engineers \cite{weisz2021perfection, weisz2022better} & 2 & 2.8\\
    & General public/Unspecified & General public/Unspecified \cite{hu2018touch, mittal2020photo, costa2018automatic, laban2022newspod, lee2018content, liu2021transformer, wallace2021learning} & 33 & 45.8\\
    & Disabled & Blind or low vision people \cite{hofmann2022maptimizer} & 1 & 1.4\\
    & Researchers & Researchers \cite{zhang2021method,  putze2018detecting}; Graduate students \cite{gero2022sparks} & 6 & 8.3\\
    & Writers & Novelists \cite{10.1145/3491102.3501819}; Amateur writers \cite{yuan2022wordcraft}; Poets \cite{10.1145/3290605.3300526} & 7 & 9.7\\
    & Etc. & Domain experts (gesture generation \cite{yoon2021sgtoolkit}, image restoration \cite{weber2020draw}) & 2 & 2.8\\
\midrule
    \multirow{8}{*}{Task} & Composition & Human-AI music co-creation \cite{louie2020novice}; Music for video generation \cite{frid2020music}; Drum beat creation \cite{vogl2019automatic} & 9 & 12.5\\
    & \textbf{Design} & \textbf{Furniture design \cite{urban2021designing}; 3D object design \cite{matejka2018dream}; Graphic design \cite{ueno2021continuous}} & \textbf{13} & \textbf{18.1}\\
    & \textbf{Drawing} & \textbf{Illustration \cite{liu2022opal}; Computational drawing \cite{wang2020latent}; Human-AI co-drawing \cite{fan2019collabdraw, oh2018lead, 10.1145/3377325.3377522}} & \textbf{10} & \textbf{13.9}\\
    & Education & Teaching programming \cite{jonsson2022cracking, suh2022leveraging} & 2 & 2.8\\
    & Everyday-work & Podcast listening \cite{laban2022newspod}; Video captioning \cite{yuksel2020human}; Conversation  \cite{huber2018emotional} & 11 & 15.3\\
    & Programming & Code generation \cite{jiang2022discovering}; Code translation \cite{weisz2021perfection, weisz2022better} & 3 & 4.2\\
    & Writing & Story writing \cite{10.1145/3532106.3533506, yuan2022wordcraft}; Email writing \cite{buschek2021impact, liu2022will}; Fictional character writing \cite{schmitt2021characterchat} & 13 & 18.1\\
    & Etc. & Facial expression generation \cite{abdrashitov2020interactive}; Gesture generation  \cite{yoon2021sgtoolkit}; Image restoration \cite{weber2020draw} & 11 & 15.3\\
\midrule
    \multirow{9}{*}{\shortstack{Role\\(multi-label)}} & Automation & The generative model is used to automate a task without any user interface. & 29 & 40.3\\
    \cmidrule{2-5}
    & \multirow{2}{*}{Co-work} & \multirow{2}{*}{\shortstack[l]{The generative model is integrated into an interactive system to help user perform\\ a specific task.}}& \multirow{2}{*}{30} & \multirow{2}{*}{41.7}\\
    & \\
    \cmidrule{2-5}
    & Exploration & The generative model is used to help user's ideation process. & 32 & 44.4\\
    \cmidrule{2-5}
    & \multirow{2}{*}{Mediation} & \multirow{2}{*}{\shortstack[l]{The generative model is used to facilitate or arbitrate in communication between \\ the persons concerned.}}& \multirow{2}{*}{10} & \multirow{2}{*}{13.9}\\
    & \\
    \cmidrule{2-5}
    & Representation & The generative model is used to embed information in a latent space for downstream tasks. & 9 & 12.5\\
\bottomrule
\end{tabular}
}
\end{center}
\end{table*}

To understand how LTGMs have been leveraged in previous research, we conducted a systematic literature review.
For this aim, we scrutinized a total of 72 system and application papers from several prominent HCI venues.

\subsection{Identification}
\label{sec3:identification}
\subsubsection{Venue.}
We examined six primary sources known for quality HCI research: the ACM CHI Conference on Human Factors in Computing Systems (CHI); the ACM Symposium on User Interface Software and Technology (UIST); the ACM Conference on Intelligent User Interfaces (IUI); the ACM SIGCHI Conference on Designing Interactive Systems (DIS); the ACM Conference on Computer-Supported Cooperative Work and Social Computing (CSCW); the ACM Conference on Creativity \& Cognition (C\&C).
AI venues were excluded since we wanted to focus on the human-side understanding of the research.
Particularly, we wanted to understand how humans can utilize technology to support their work, rather than quantitative evaluation or advancement of state-of-the-art AI technologies.
We constrained the publication period to the past five years (i.e., from 2018 to 2022) to reflect recent trend of the research \cite{speicher2019mixed}.

\subsubsection{Search Keyword.}
Since it has been just a year since DALL-E was released, there exists only a handful of HCI research that focused on LTGMs directly.
Instead, we investigated papers on generative models, considering that LTGM is another type of generative model trained at scale \cite{bommasani2021opportunities}.
For a thorough inspection of all the relevant publications in the above venues, we tried to be inclusive as much as possible.
Namely, rather than finding papers that contained the exact term `generative model', we tried to embrace all existing papers which include extended forms of search keywords such as `generate', `generation', `generative', and `generating' in their title or abstract.

\subsubsection{Exclusion Criteria.}
After running an exhaustive search, we set 6 exclusion criteria, referencing previous notions \cite{scuri2022hitting, baykal2020collaborative}, to eliminate papers that do not align with the goal of this research.

\begin{itemize}
  \item [EC 1:] The common use of the term `generate', `generation', `generative', and `generating' do not align with our research direction. For example, we removed if the term `generative’ was used as a general adjective, not used to mean generative models. This criteria was required as we gathered papers inclusively in round 1.
  \item [EC 2:] We chose system/application papers that used generative models to support their tasks. In this regard, qualitative/study papers \cite{oh2020understanding, ross2021evaluating} were excluded.
  \item [EC 3:] We chose papers that use or study neural network-based generative models such as GAN, VAE, LSTM, GPT-3, and DALL-E. For instance, the use of conventional generative models like Gaussian mixture models was excluded.
  \item [EC 4:] Since we are interested in the impact of generative models to humans, excluded are the technical papers which adopted generative models for data augmentation \cite{maghoumi2021deepnag, li2021crossgr}, synthesizing \cite{janveja2020insight}, and complement \cite{wang2021thru}.
  \item [EC 5:] We excluded doctoral dissertation, survey paper, book, book chapter, and demo papers.
  \item [EC 6:] Lastly, we excluded the paper published by the same author(s) where the content is not different significantly.
\end{itemize}

\subsubsection{Round 1---4.}
Our strategy to perform the literature review was to collect papers broadly and then narrow down the scope for careful examination \cite{10.1145/3290605.3300619}.
Therefore, in round 1, we collected a total of 1,494 papers from the 6 venues (i.e., CHI, UIST, IUI, DIS, CSCW, C\&C) using the search terms (i.e., generate, generation, generative, and generating) on a popular academic database, ACM Digital Library\footnote{https://dl.acm.org}.
In round 2, papers with the exact same title and content were removed (i.e., EC 6), which left 1,429 papers.
In round 3, we read the title and abstract of the remained ones, and checked EC 1---5 to exclude papers that do not fit in the goal of our research.
Lastly, in round 4, we read all the articles that passed all the previous rounds (i.e., 64 papers), and additionally added in 8 papers that were not found during the above process.
Thus, we finalized our collection as a set of 72 papers.
\autoref{tab:review-number} shows the precise number of papers in each round.


\subsection{Analysis Procedure}
Grounded on previous survey papers \cite{10.1145/3290605.3300619, disalvo2010mapping, baytas2019design}, we built a question rubric to investigate each paper on a deeper level.
We tried to answer the following questions while reading the selected papers: 1) What are the roles of the generative model in the research?
2) What are the research questions? If they are not explicitly stated, what are the main contributions of the research?
3) Who are the target users of the system/application? What are the tasks given to them in evaluating their system/application?
4) What are the target users' opinions about the system/application?
5) What are the discussions and takeaway messages of the research?

For the analysis of the identified publications, we performed an open coding on the usage of generative models.
To start with, the first author answered the question rubric on each paper.
Based on the answers, two authors read it multiple times to be familiar with the data to generate the initial low-level codes, independently.
Afterwards, they revised the codes collaboratively, iterating three times until the final high-level codes were determined.
Next, four authors clustered related codes and searched for the themes that could embrace the overall literature.
Finally, they decided the names of each theme that represented its codes well.

\subsection{Results}
\label{sec3:result}

We identified three themes of the publications: user (i.e., the target group of the research) containing 9 codes, task (i.e., the main task suggested in the research) including 8 codes, and role (i.e., the purpose generative model is used in the research) with 5 codes.
The detailed information is denoted in \autoref{tab:review-utr}.

\subsubsection{User.}
We noticed that generative models were used to help diverse types of users.
In total, we found 9 codes: \textbf{artists}, children/students, \textbf{designers}, engineers, general public/unspecified, the disabled, researchers, writers, and etc.
The vast majority of the publications targeted general public or they simply unspecified the user (45.8\%, 33 publications).
Next, \textbf{13 publications (18.1\%)} targeted \textbf{designer} as their user.
Different types of designers were included throughout the papers including fashion designers \cite{padiyath2021desainer}, graphic designers \cite{ueno2021continuous}, poster designers \cite{guo2021vinci}, and so on.
We found 7 publications (9.7\%) which targeted writers like novelists \cite{10.1145/3491102.3501819}, amateur writers \cite{yuan2022wordcraft}, and poets \cite{10.1145/3290605.3300526}.
There were 6 publications (8.3\%) that targeted researchers such as AI researchers, UX researchers, and graduate students.
We identified \textbf{5 publications (6.9\%)} which targeted \textbf{artists} like digital artists \cite{yurman2022drawing}, 3D modelers \cite{abdrashitov2020interactive}, and webtoon authors \cite{wetoon}.
There were 3, 2, and 1 publications (4.2\%, 2.8\%, and 1.4\%) for children/students, engineers, and the disabled, respectively.
Lastly, 2 publications (2.8\%) were classified as etc. which includes domain experts in gesture generation \cite{yoon2021sgtoolkit} and image restoration \cite{weber2020draw}.

\subsubsection{Task.}
A total of 8 codes were identified in the review process.
Specifically, writing was the main task in 13 publications (18.1\%).
This includes different writing scenarios such as story writing \cite{10.1145/3532106.3533506, yuan2022wordcraft}, email writing \cite{buschek2021impact, liu2022will}, and scientific writing \cite{gero2022sparks}.
Everyday-work was the target in 11 publications (15.3\%) such as podcast listening \cite{laban2022newspod}, video captioning \cite{yuksel2020human}, and conversation \cite{huber2018emotional} which can be encountered in our everyday life.
There were \textbf{13 publications (18.1\%)} targeting \textbf{design}, which captured various sub-domains such as fashion design \cite{padiyath2021desainer}, furniture design \cite{urban2021designing}, UI/UX design  \cite{gajjar2021akin, ang2021learning, liu2018learning, wang2021screen2words, mozaffari2022ganspiration}, and graphic design \cite{ueno2021continuous}.
We identified \textbf{10 publications (13.9\%)} which targeted \textbf{drawing} including illustration \cite{liu2022opal}, computational drawing \cite{wang2020latent}, human-AI co-drawing \cite{fan2019collabdraw, oh2018lead, 10.1145/3377325.3377522}, and webtoon drawing \cite{wetoon}.
There were 9 (12.5\%), 3 (4.2\%), and 2 (2.8\%) publications in composition, programming, and education, respectively.
Lastly, 11 publications were grouped as etc. (15.3\%) which contained tasks like tactile map generation \cite{hofmann2022maptimizer} and choosing photo-realistic images \cite{zhang2021method}.

\subsubsection{Role.}

We recognized five codes as the role of the generative models, which are automation, co-work, exploration, mediation, and representation.
Unlike other themes, multiple codes were allocated to a single publication if needed, since we found it inefficient to build the structure to become mutually exclusive and exhaustive at the same time.
However, no more than two codes were simultaneously assigned to a single paper.
Our codes came out differently from a previous work that delved into creativity support tools \cite{chung2021intersection}, likely because we focused only on papers with generative models.


The automation means that the generative model was used to automate or replace human efforts to perform a specific task without any user interface (40.3\%, 29 publications).
For example, Singh et al. \cite{mittal2020photo} introduced a prototype to create a personalized emoji from a rough drawing where the user's level of intervention was extremely restricted.
The co-work indicates that the generative model was integrated into an interactive system to help perform a task in a human-in-the-loop or a mixed-initiative approach (41.7\%, 30 publications). 
For example, \textit{Calliope} \cite{bougueng2022calliope} provided a user interface containing several levels of granularity to manipulate parameters (e.g., tempo) in generating MIDI files.
Although both automation and co-work support the users in doing their tasks, the main difference lies in whether there exists any user interface with which humans can interfere.

The exploration signifies that the generative model was used to help users' creative thinking process so they could get inspired to produce novel ideas (44.4\%, 32 publications).
For instance, Fede et al. presented \textit{The Idea Machine} \cite{di2022idea} using large language models to empower user's idea generation in writing.
The mediation means that the generative model was used to facilitate or arbitrating in communication between the concerned parties (13.9\%, 10 publications).
\textit{EmoBalloon} \cite{aoki2022emoballoon} used generative models to make emotional speech balloons to decrease the gap of emotional arousal between the sender and receiver of a text message.
Lastly, the representation means that the generative model was used to embed information in a latent space for downstream tasks (12.5\%, 9 publications).
For example, \textit{Screen2Vec} \cite{li2021screen2vec} showed semantic embeddings of GUI screens and components.
\begin{table*}[t]
\caption{\textbf{Detailed background information of 28 interview participants.}
A total of 28 professional visual artists with varying occupations were recruited via multiple mediums.
They covered 35 domains as some have expertise in multiple fields.
Their years of experience ranged from 1 to 19 years.}
\label{tab:interview}
\begin{center}
\resizebox{\linewidth}{!}{%
\def\arraystretch{1.25}
\begin{tabular}{lllll}
\toprule
    & Current Occupation & Experience & Major & Examples of Created Artifacts \\
    \hline
    \grayrow
    P1 & Webtoon author & 10 years & Metal craft design & Cartoons published on the web\\
    
    P2 & Video editor & 3 years & Mass communication \& journalism & YouTube videos\\
    \grayrow
    P3 & Web/UI/UX designer & 6 years & Visual communication design & UI/UX design of a medical app\\
    P4 & Motion graphic designer & 3 years & Visual communication design & 2D and 3D motion graphics\\
    \grayrow
    P5 & Product designer (Mobility app) & 4 years & Service and design engineering & Product design of a mobility app\\
    \multirow{2}{*}{P6} & Illustrator & 4 years & \multirow{2}{*}{Painting} & Album covers; Concept art\\
    & Picture book writer & 2 years & & Picture books for young children  \\
    \grayrow
    P7 & Artist (Ceramics) & 4 years & Sculpture & Ceramic art for exhibitions\\
    P8 & Product designer (Medical app) & 7 years & Visual communication design, painting & UI/UX design of a medical app\\
\grayrow
    P9 & Jewelry designer & 4 years & Metal craft design & Commissioned jewelry design\\
    P10 & Artist (Abstract art) & 6 years & Fine arts & Abstract painting for exhibitions\\
\grayrow
    & Web/UI/UX designer & 1 years & & UI/UX design of a medical app\\
\grayrow
    \multirow{-2}{*}{P11} & Spatial designer & 2 years &\multirow{-2}{*}{Interior architecture design} & Visual merchandising\\
    \multirow{2}{*}{P12} & \multirow{2}{*}{Artist (Video, painting, sculpture)} & \multirow{2}{*}{5 years} & \multirow{2}{*}{Fine arts} & \multirow{2}{*}{\shortstack[l]{Metalworking, painting, and filming\\ for exhibitions}}\\
    \\
\grayrow
    P13 & 3D character modeler (Game) & 4 years & Game media & 3D character modelling\\
    \multirow{2}{*}{P14} & Product designer (Character animation) & 8 years & \multirow{2}{*}{Animation design}& Product design of toys for children\\
    & Amateur cartoonist & 1 years & & Cartoons published on Instagram\\
\grayrow
    & Graphic designer & 1 years & & Motion graphics for lectures\\
\grayrow
    & Art instructor & 3 years & & Art class for elementary school students\\
\grayrow
    \multirow{-3}{*}{P15}& Webtoon author & 3 years & \multirow{-3}{*}{Oriental painting, fine arts}& Cartoons published on the web\\
    P16 & Concept artist (Movie, game) & 12 years & Digital animation & Concept art for movies and games\\
\grayrow
    P17 & Artist (Video, painting) & 2 years & Fine arts & Painting and filming for exhibitions\\
    P18 & Fashion designer & 4 years & Fashion design & Fashion design for jackets, shirts, and pants\\
\grayrow
     & Artist (Movie, photography) & 6 years & 
    & Photography and filming for exhibitions\\
\grayrow
    \multirow{-2}{*}{P19} & Photographer & 2 years & \multirow{-2}{*}{Fine arts} & Photos for art brochure\\
    P20 & Industrial designer & 3 years & Industrial design & Design for furniture and electronic devices\\
\grayrow
    P21 & Artist (Sculpture) & 7 years & Sculpture & Sculpture for exhibitions\\
    \multirow{2}{*}{P22} & Editorial designer&  5 years & \multirow{2}{*}{Character design} & Commissioned poster design\\
     & Logo designer & 2 years & & Commissioned signboard design\\
\grayrow
    & Art instructor & 5 years & & Art class for adults and children\\
\grayrow
     & Art program director & 3 years & & Art program planning\\
\grayrow
    \multirow{-3}{*}{P23}  & Artist (Contemporary art) & 5 years & \multirow{- 3}{*}{Korean painting}& Painting for exhibitions\\
    P24 & Projection designer & 6 years & Film (minor in Multimedia)& Projection design for theatrical plays\\
\grayrow
    P25 & VFX artist (Game) & 4 years & Media content & Casual game effects\\
    P26 & Artist (Painting, video) & 19 years & Painting & Painting and filming for exhibitions\\
\grayrow
    P27 & Architect & 5 years & Architecture & Building design\\
    P28 & 3D animator & 12 years & Digital content & 3D movie animation\\
\bottomrule
\end{tabular}
}
\end{center}
\end{table*}

\section{Visual Artist Interviews}
\label{sec4}

Based on the findings of the systematic literature review, we performed interviews with 28 professional visual artists with various occupations to answer the following research questions:

\begin{itemize}
  \item \textbf{RQ1.} Which subgroups of visual artists would be more willing to use LTGMs?
  \item \textbf{RQ2.} What types of tasks would they utilize LTGMs on?
  \item \textbf{RQ3.} What would the LTGMs' role be in those cases?
\end{itemize}

\subsection{Participants}

We defined the visual art as \textit{the art expressed by visual elements}, and define visual artist as those who work in visual art domains.
Following the definition, visual art includes extensive fields like painting, sculpture, ceramics, photography, film, multimedia, design, and craft. Thus, we tried to contain various occupations in recruiting them.
As shown in \autoref{tab:interview}, a total of 28 visual artists participated in the interview.
The interview participants were recruited via diverse mediums such as professional networking and career development apps (e.g., Linkedin\footnote{https://linkedin.com}), social networking services to share photos and videos (e.g., Instagram\footnote{https://www.instagram.com}), and platforms for visual artists' self-promotion to showcase their artworks (e.g., Artstation\footnote{https://artstation.com}).
We also used the snowball sampling approach for additional recruitment \cite{goodman1961snowball}.

The participants covered 35 unique visual art domains as some participants possessed expertise in multiple fields.
We did not filter out the interview participants in advance except for two constraints that 1) they have work experiences of at least a year in their fields, and 2) they have never experienced LTGMs before the interview to prevent any preconception \cite{yuan2022wordcraft}.
The participants' average work experience was 6.1 years, but the years of experience in their fields was higher as most of them received higher education in the relevant fields.
Each participant was provided 20,000 KRW as a compensation.


\subsection{Interview Protocol}

We proceeded with semi-structured interviews on a remote condition which lasted 67 minutes on average with a minimum of 52 minutes to a maximum of 88 minutes.
Both the video and audio were recorded with consent and transcribed into text for a thorough analysis.
The interview consisted of three parts: 1) asking questions about the participants' demographics and their domain, 2) experiencing LTGMs, and 3) eliciting participants' thoughts on how to use LTGMs for their creative works.
In the second part, we used a specific LTGM (i.e., DALL-E).
However, we induced the participants to focus on its potential functionalities, rather than what it can and cannot do at the moment, as we did not want to confine the study result under the current limitations of DALL-E.
After the participants signed the consent forms, we asked several questions regarding their domains to understand their working process in detail.
The questions are listed below:
\begin{itemize}
  \item Could you describe your job?
  \item What are the typical tasks you perform?
  \item Are there specific tasks that are unnecessarily repetitive or that require a lot of creativity?
  \item Where do you get inspirations, if any?
  \item Do you use any references for work? How do you use them?
  \item How much communication and collaboration do you encounter at work? What are the difficulties with them?
\end{itemize}

In addition, we asked 7-point Likert scale questions about their familiarity (1=not at all familiar, 7=extremely familiar) and personal attitude toward AI (1=very negative, 7=very positive) to take account of previous notions on participants' behavior based on preconception \cite{hsu2021attitudes, milanovic2021misattribution}.
To introduce DALL-E's image generation performance over varying fields, we showed them a lot of text prompts and corresponding image samples generated by DALL-E in several domains (e.g., architecture, fashion, fine art, industrial design, UI/UX, etc).
After the demonstration, participants were given sufficient time to use DALL-E's three functionalities: image generation, variation, and synthesis.
We requested them to follow a think-aloud procedure~\cite{van1994think}.
The interview was finished by asking final interview questions:
\begin{itemize}
  \item How would you adopt LTGMs for creative works in your domain?
  \item What are the differences between LTGMs and previous tools?
  \item What would be the works that cannot be supported by LTGMs?
  \item Are there any additional functionalities you want?
  \item How would LTGMs change the working paradigm in your domain?
\end{itemize}


\subsection{Analysis Method}


For the analysis of the transcribed data, we firstly deployed a deductive coding framework with the themes (e.g., User, Task, and Role) drawn in \autoref{sec3:result}.
Afterwards, we performed an inductive coding \cite{braun2006using} to investigate additional codes, which were iteratively refined through discussions until consensus was reached.
In the overall procedure, we used a qualitative data analysis tool, namely Dovetail~\cite{dovetail}, to organize and aggregate relevant data.
Specifically, the first and the second authors read all the transcriptions, and labeled sentences with existing codes, independently, which resulted in an initial codebook regarding the pre-built themes.
Since previous themes and codes were developed upon the use of generative models and all types of users, we re-examined the codebook inductively to generate an updated one specific to our research question (i.e., visual artist and LTGMs).
We iterated the revision by merging and dividing the codes multiple times until we unanimously agreed to achieve a theoretical saturation \cite{sandelowski1995sample}.
Since all interviews were conducted in Korean, we created codes in the same language to prevent possible change of nuance and loss of certain meanings in translation \cite{van2010language}.


\section{Findings}
\label{sec5}

We introduce the potential and limitations of LTGMs.
To report possible use cases, we categorized our interview findings based on three themes.
Since our research focused mainly on visual artists, we had to modify our codebook. 
In the literature review, we only had two codes in `User' (i.e., artist and designer) and two codes in `Task' (i.e., drawing and design) related to the visual art domain, but we subdivided them into more detailed characteristics.
We also present four limitations that LTGMs cannot properly support the visual artists in their work flows.







\subsection{Potential of LTGMs}

\subsubsection{Image Reference Search Tool}
Among 28 interviewees, 12 acknowledged that LTGMs can be a new image reference search tool.
They pointed out that referencing visual materials (e.g., images and videos) is a crucial part of their creative work.
They mainly use references to 1) learn by observing what and how others create and 2) get inspirations for new ideas.
In general, it helps to realize their imagination into the world.
However, there were some exceptions as well.
For instance, P21 (sculptor) uses references to check that the work she plans to make does not yet exist.
References that are irrelevant to their field of domain are also utilized by 7 visual artists, suggesting that they help eliciting novel ideas.
P20 (industrial designer) said:
\begin{quote} 
\textit{\textbf{P20, Industrial designer:} If I were to design a speaker, I do not just find a bunch of speaker images. Rather, I find various images ranging from fine arts to architecture, and gather them all together as a collection. Next, the images are classified into several themes following the client's design requirements to make several design prototypes.}
\end{quote}

All the reference seekers complimented that LTGMs are fast, convenient, and can make a large number of high-quality images that are unique and different from one another.
Specifically, P9 (jewelry designer) acknowledged that LTGMs seem to be useful in the early stages like planning where a huge amount of images are needed for ideation.
Also, P23 (contemporary art artist) argued that she felt more passive when using Google, but felt more active with LTGMs since they gave more curated samples using functionalities that require more human engagement (e.g., selecting regions with brushing).
Despite all the aforementioned advantages and distinct features, 10 of the 12 reference seekers said LTGMs would not become a tool to shift their working paradigm, rather just another option for reference searching.
P25 (VFX artist) stated:
\begin{quote}
\textit{\textbf{P25, VFX artist:} The advantage (of LTGMs) is that it can generate any images with several detailed conditions. Although Google is big, it cannot provide if the image is not contained in its database... Frankly speaking, I think people will use LTGMs, but there would not be a huge difference. Currently, if there are 100 people, around 90 use Pinterest. However, in the future, some of them will say `I use LTGMs'. That is all.}
\end{quote}




The detailed codes of User, Task, and Role are the following:
\begin{itemize}
  \item User: Visual reference seeker
  \item Task: Ideation
  \item Role: Exploration
\end{itemize}

\subsubsection{Enabling Fast Real-time Visual Communication}

Through the interviews, we confirmed that 20 visual artists acknowledged LTGMs would be beneficial in real-time visual communication for those who 1) work in a company adopting a top-down decision-making procedure, 2) commission to provide professional services to clients, and 3) interact with varying artists in different fields.
Specifically, 11 visual artists who communicate with their boss back and forth in verifying and improving their work wanted to leverage LTGMs for a fast prototyping.
They mentioned that the concept of art delivered from their boss is often vague, which requires several exploration in varying directions, especially in early stages.
Therefore, they have to make many drafts as alternatives for confirmation before finalizing the concept, with most of the drafts ending up abandoned.
The visual artists want to save their time and energy by taking advantage of LTGMs for such dissipated work.
For example, P1 (webtoon author) stated that using LTGMs could decrease the time consumed in communication between her and her manager by performing a real-time revision of the high-level concepts.
Specifically, she said she wants to make several images and let the manager make the concept as an ensemble by extracting and combing parts from the generated images.
Similarly, P16 (concept artist) pointed out that LTGMs could guide the working direction in a case where the boss does not have a concrete goal in mind.
P16 said:
\begin{quote}
\textit{\textbf{P16, Concept artist (movie, game):} While some art directors indicate a specific direction, others do not have one, but just want me to give some ideas. In this case, I have to show them many drafts to see their reaction and decide what to focus on and what not to... We usually work as a group of concept artists, and the director want us to draw the concept art with the same topic, respectively. If one is picked, then it is branched out further. We could use LTGMs to generate starting pieces for conversation in the initial discussion we can work based upon.}
\end{quote}


Moreover, 14 visual artists mentioned that LTGMs would help facilitate communication between the visual artists and their clients.
Many visual artists commission artworks, offering professional services.
In doing so, they have to meet the client's requirements, which can often be ambiguous and even paradoxical at times.
P8 (product designer) mentioned that once a client wanted a design that has both a `cold' and `warm' mood at the same time.
She said the best way to communicate with the clients is often using visual materials where LTGMs can be a viable means.
Similarly, P4 (motion graphic designer) stated that LTGMs would make it much easier to persuade clients as it enables fast prototyping.
She added that communication often requires an iterative process, but she would be able to create a bunch of rough images in a short time using LTGMs.
Moreover, when collaborating with many coworkers, it is essential to narrow down the gap between their thoughts on the final artifact.
P14 (product designer) said LTGMs could decrease the time needed to reach a consensus:
\begin{quote}
\textit{\textbf{P14, Product designer (character animation):} Our work requires communication (between workers) as mandatory, as the design of a product can be pleasing only to me (but not to others). By taking others' feedback, we calibrate the direction on how to revise it. Thus, we try to take comments as much as we can before embarking on real work. LTGMs can decrease the time for the overall communication process.}
\end{quote}






Lastly, 2 visual artists mentioned that LTGMs could be useful for those who interact with people across multiple domains.
In detail, P24 (projection designer) pointed out that the communications with dancers and musicians are different as it requires a domain-specific understanding.
He ascribed the fundamental reason to the inconsistent connection between text and visual materials:
\begin{quote}
\textit{\textbf{P24, Projection designer:} The language used by dancers, projection designers, and musicians is all very different. For example, dancers express abstract words, such as `joy' through their work, but projection designers often find it hard to do the same with videos (as they do not have a logical connection). Therefore, we communicate via a bunch of images and ask them which one is visually close to your thoughts on `joy', for example. LTGMs can be of use in such cases to narrow down (the search space) and 
reach an agreement.}
\end{quote}

The detailed codes of User, Task, and Role are the following:
\begin{itemize}
  \item User: Visual communicator; Client/Contractor in visual art work
  \item Task: Real-time visual communication
  \item Role: Mediation
\end{itemize}



\subsubsection{Rectifying Human's Biased Creation}

In the study, 14 visual artists said that LTGMs could help them try unconventional things and think out of the box.
Interestingly, 8 visual artists directly mentioned that humans have a bias in creating art, but LTGMs do not seem to have one.
This is a finding that contradicts a previous research which revealed that people have a preconception on AI that it generates biased artifact \cite{bennett2021s}.
The visual artists pointed out that humans have preferences, which can lead them to certain directions in performing creative works.
They wanted to use LTGMs for attempting what they have never tried, deviated from their comfort zones.
P17 (artist) said:
\begin{quote}
\textit{\textbf{P17, Artist (video, painting):} The advantage of LTGMs is that they do not have a bias (in generating art). Thus, it can generate any combination I make in a totally unexpected way. My personal knowledge and skills will be reflected to the work in previous approach, for example, the working style I stick to so far. However, the output of LTGMs is random, which is good in that it can recommend a new style of work.}
\end{quote}

Aside from personal preferences, 3 out of 8 visual artists mentioned that there are more practical reasons that prevents visual artists from taking unusual approaches.
They said that it necessitates a lot of time and energy but often results in an unsatisfactory outcome.
However, they praised that LTGMs encourage visual artists to take adventurous approaches, as it shows what they have in mind in a very small amount of time.
P6 (illustrator and picture book artist) said:
\begin{quote}
\textit{\textbf{P6, Illustrator and picture book artist:} Although I desire to draw a night sky with purple color and a yellow house (as I think they look good overall), in reality, they might not harmonize with each other. Also, I need to think about the size of house and all the other details... Until the best one is picked, it takes a lot of time. Along with such a long procedure, I cannot find most of what I wanted at the first place from the final result. LTGMs can be a viable solution to make my imagination to real in short time.}
\end{quote}

The detailed codes of User, Task, and Role are the following:
\begin{itemize}
  \item User: Unconventional visual artist
  \item Task: Unbiased prototyping
  \item Role: Automation
\end{itemize}





\subsubsection{Low-fidelity Prototyping for Novice Visual Artists}

A total of 9 visual artists pointed out that it would be helpful for those 1) who have low expertise in handling domain-specific art creation tools, and 2) who have little knowledge in related art fields, because LTGMs can help them generate low-fidelity prototypes.
Since new software comes out every year, visual artists need to learn new skills constantly to not fall behind.
However, the novices have difficulties in knowledge acquisition because it takes huge amount of time.
For example, P7 (artist) did not use any software but performed manually for making a blueprint of artwork because she was too busy during working hours, although she knew it was inefficient.
She wanted to leverage LTGMs in her work, as they can handle time-consuming tasks like image synthesis using simple interactions (e.g., brushing) and text prompts.
Also, they said working in the visual art domain often requires knowledge and competencies across multiple fields.
For example, since the company which P4 (motion graphic designer) works for only had two designers, she had to undertake all design-related tasks, although she only had experience in 2D motion graphics.
P4 stated that LTGMs can help visual artists to embrace tasks beyond their capacity.
Although the quality of created prototypes would be low, it would be of great help, when the outcome is needed urgently.

The detailed codes of User, Task, and Role are the following:
\begin{itemize}
  \item User: Novice visual artist
  \item Task: Low-fidelity prototyping
  \item Role: Automation
\end{itemize}

\subsubsection{Justification Tool for "a New Era of AI Art"}


We found that LTGMs could be a tool to create artworks as well as justification for them in the fine art domains.
Specifically, P12 (artist) wanted to use LTGMs in creating his painting because it is more credible once AI guarantees the generated results on a subjective concepts.
Also, he mentioned artists prefer to take advantage of new technologies, so there can arise a new artistic paradigm called "an era of AI art" that is a complete prosperity of AI-generated contents:
\begin{quote}
    \textit{\textbf{P12, Artist (video, painting, sculpture):} As my topic of art is ugly people, I would like to generate 100 couples of men and women using LTGMs. Then, I would synthesize each pair to generate 50 pairs as their descendants. Through iterations, there would be the final one who is the epitome of an ugly person.}\\
    \textit{\textbf{Interviewer:} We can do that using Google. What is different from using LTGMs in doing that work?}\\
    \textit{\textbf{P12, Artist (video, painting, sculpture):} It is hard to find many different images for a certain text prompt in Google. Moreover, if I find images of ugly person on Google, others might disagree as it is based on subjective perception. However, I can justify my claim on the artwork if the generated images are made by AI. Who would argue with that if AI says so?}
\end{quote}

The detailed codes of User, Task, and Role are the following:
\begin{itemize}
  \item User: Unconventional artist
  \item Task: Creating and justifying generated artworks
  \item Role: Automation
\end{itemize}



\begin{figure*}[t]
  \centering
  \includegraphics[width=\linewidth]{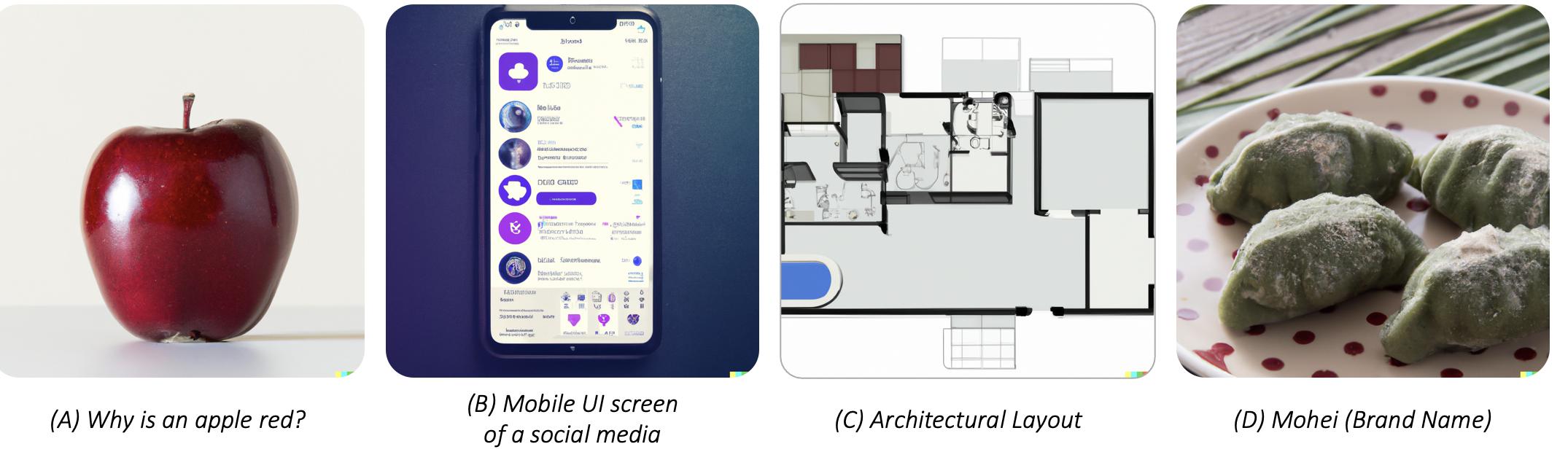}
  \caption{\textbf{Limitations of LTGMs.} (A) Although P8 wanted an image that enables philosophical reasoning, the output was an image of a red apple. (B) P3 thought the UI screen does not reflect the user experience, thus the design is not practically useful. (C) P27 noticed that the layout does not consider practicality so it cannot be built by humans. (D) P8 tried to type in her brand name, `Mohei' which means the sunshine on the corner, but the output was a totally irrelevant image. }
  \label{fig:limitation}
\end{figure*}

\subsection{Limitations of LTGMs}
\label{sec5:limit}

\subsubsection{LTGMs only generate predictable images}

A total of 5 visual artists said LTGMs are logical machines that can only create predictable images, when what they wanted were random explorations.
Because getting inspiration comes from unexpected events, they wanted to find images that do not exactly match with their input text prompt.
P22 (editorial designer and logo designer) mentioned that there exists a logical connection between the images generated by LTGMs and the input text prompt, making it impossible to elicit any novel ideas further.
Moreover, P11 (spatial designer) mentioned that it is important to add such randomness in work, as it looks tedious and outdated if the concept is used directly.
P11 said:
\begin{quote}
    \textit{\textbf{P11, Spatial designer:} We work based on the concept and clothing to display. Then, we plan how to decorate the place considering the fabric and the atmosphere of the season... Although our concept is a flower road, for example, we cannot just display a flower road; that feels old-fashioned... That is why we find references from diverse sources.}
\end{quote}

Moreover, they pointed out that LTGMs cannot create images that embody philosophical meanings, storytelling, or sophisticated interpretation.
For instance, P8 (product designer), who studied both fine arts (painting) and design, mentioned that the images generated by LTGMs are not interesting from the painter's perspective.
She was even able to predict what the outcome image of LTGMs would be in advance.
In specific, she said LTGMs would create an ordinary red apple when she type in `why is an apple red?', and the outcome was the same as her expectation (\autoref{fig:limitation}-A).
However, what she wanted was an image conceptualizing the philosophical reasoning that people can take to a further contemplation.

\subsubsection{LTGMs Do Not Support Personalization}

A total of 11 visual artists doubted whether LTGMs could generate images that require domain-specific understanding.
For example, visual artists thought LTGMs are hard to reflect user experience and practicality.
For example, P3 (Web/UI/UX designer) was surprised to see the generated image of mobile UI screen, but was uncertain about LTGMs' capabilities to consider the user flow.
He worried that it will just visualize the given conditions without considering the user experience (\autoref{fig:limitation}-B).
P18 (fashion designer) and P27 (architect) gave similar opinions.
They acknowledged that LTGMs can generate rough sketches of artifacts (i.e., clothing and building), but it cannot handle all the detailed requirements of the design.
Specifically, P27 mentioned that architects have to consider practical issues like whether it can actually be constructed by humans while designing the architectural layout.
However, she thought LTGMs would not include realistic considerations in the layouts (\autoref{fig:limitation}-C).

Moreover, the visual artists wanted to incorporate their identities into the artifacts.
For example, P10 (artist) stated how visual artists would not be able to add their own identity to the artifact if LTGMs managed all the work.
Similarly, P17 (artist) wished for customization of the generated images to reflect her own style since she worried that they would contain some sort of characteristics of LTGMs, so that they do not seem something special, but stale artwork.

\subsubsection{Text Prompting Restrains Creativity}
A total of 9 visual artists pointed out that LTGMs cannot generate novel images due to their dependence on text prompting.
We found two main reasons for this.
First, since LTGMs use text prompts, they needed to have a word or phrase to describe what users want to create.
However, when creating something that is entirely new, there would be no word for it yet, which results in a paradoxical situation.
For example, P8 (product designer) who designs a medical app tried to generate images using the text prompt `Mohei', which is a provisional brand name for her own product.
The original meaning of Mohei was an abbreviation of `sunshine on the corner', but the output image was a totally random image such as a dumpling (\autoref{fig:limitation}-D).
Similarly, P10 (artist) who draws abstract art mentioned that it is impossible to find pertinent text prompt for realizing his thought.
This is because he prefers to seek a possibility that comes from consecutive random brush strokes, which is an artistic technique called impasto.
His art becomes completed through a ceaseless interaction with the canvas.
Therefore, he found it is hard to describe the overall art-making process in texts.
P10 stated:
\begin{quote}
    \textit{\textbf{P10, Artist (abstract art):} I worry about the fundamental limitation of text prompting. In the planning, I do not use text, but find what I want via painting. For instance, a word `dark' can represent several imagery with varying levels and combinations of darkness. There can be subtle differences between them, and I love to focus on the gap from which I can discover such representation.}\\
    \textit{\textbf{Interviewer:} What is the problem of describing it into a text?}\\
    \textit{\textbf{P10, Artist (abstract art):} The delicate feeling will volatilize while translating it. As I need to take time to choose which one is a right word, the first imagery came to my mind will go away.}
\end{quote}




Second, P27 (architect) mentioned that it would be really convenient if all the existing styles are trained on the LTGMs so they can create any of them using a simple text prompt.
However, she also admitted that it would limit the styles of generated images within the boundary of trained samples.
Thus, LTGMs would rather operate as something confining the visual artists' imagination.
P27 said:
\begin{quote}
    \textit{\textbf{P27, Architect:} It would be really convenient if all existing floor plans are trained on the LTGMs, and they can generate whatever I want... However, if that is possible, I believe they can rather confine the human's imagination. This is because images that can be generated by LTGMs will be considered as a set of templates of all existing floor plans that we can invent, although it is not true.}
\end{quote}


\subsubsection{LTGMs are Inefficient and Become a Burden}

A total of 6 visual artists left worrisome comments that learning how to use LTGMs can become a burden.
In the study, LTGMs often generated images somewhat differently from the visual artists' intention, so they could hardly build the mental model of LTGMs in a short time.
After a few trials of LTGMs, P11 (spatial designer) pointed out with a disappointed tone that some visual artists would prefer to draw by themselves or use familiar softwares.
Moreover, as suggested by previous work, it is often hard to find a proper text prompt because of the open-ended nature \cite{liu2022design}.
For example, P9 (Jewelry designer) and P15 (art instructor) did not know what to type in and contemplated for several minutes to write down the text prompt.

A total of 10 visual artists regarded LTGMs as inefficient in the situations where they already had a specific goal and requirements in their mind.
P18 (fashion designer) said if she wants to make pants with several design specifications, using LTGMs would be an inefficient use of time, as it needs a lot of additional time to describe them in sentences, although it cannot guarantee a satisfactory output that exists vividly in her mind.
Similarly, P4 (motion graphic designer) said it is often difficult to come up with a precise word (e.g., Baroque architecture) for what she wants, although she has the image (e.g., architectural style lavishly decorated with ornament) in her mind.
Last but not least, P28 (animator) mentioned that it is sometimes close to impossible to describe the imagery in words.
He said it is important to reflect what cannot be seen in the visual.
P28 stated:
\begin{quote}
    \textit{\textbf{P28, Animator:} Imagine that I make a girl who is ten years old and grown up as an orphan. Then we may come up with a certain mood of the girl. She might wear a shirt with a low chroma and always carry a cuddly toy around her, which are the things we can infer from her background information... As you know, these kinds of characters are created and developed over a hundred lines of stories. I can hardly think of making a new character with just a few sentences.}
\end{quote}


\section{Design Guidelines}
\label{sec6}

As Shneiderman said, a well-designed user interface can ease the burden of people's lives \cite{shneiderman2010designing}.
While having the interviews with 28 visual artists, we found that current forms of LTGMs such as DALL-E remain powerful but limited tools because of the lack of an intelligent user interface that can boost its performance.
Therefore, we suggest four design guidelines that future HCI researchers can refer to in building interactive systems leveraging LTGMs.
We prepared the guidelines to cover the limitations mentioned \autoref{sec5:limit}.

\subsection{Variability Level Specification for Different Types of Visual Art}

We propose that the variability level should be adjusted based on the visual artists' motivation.
This is because the type of images they want can highly vary depending on their objectives.
Some visual artists may want to find reference images that look totally irrelevant to the given text input, while others may want to see the ones most related to the given text.
We suggest three levels of specifications---Lookup, Inspiration, and Reinterpretation---based on their motivation and objectives for doing creative works.
First, Lookup is providing the most relevant images to a given text input.
This is what current LTGMs do.
Next, Inspiration is offering related and unrelated images at the same time.
Last, Reinterpretation is constructed with images that require a long time to comprehend their hidden connections to the text prompt.

Although it does not fit to all cases, in general, we recommend Lookup for those who work in applied art domains (e.g., design).
This is because visual artists in applied art domains have a higher chance to communicate with other people (e.g., client, collaborator, and boss), which necessitates a logical connection between the image and the text.
On the other hand, we propose Reinterpretation for the visual artists in fine art domain as it is more important for them to ponder upon the inside of themselves than communicate with other people, which results in such logical connection less important.
Thus it would be useful to have unexpected images as they can be the source for contemplation on the subject.







\subsection{Model Customization Grounded in Domain-specific Understanding}

We propose that there should be model customization to satisfy visual artists in varying domains.
Depending on the domain, each participant would have different priorities and preferences in the generated images.
For example, to generate images containing philosophical meanings, we can use a dataset of contemporary art images to fine-tune the model.
As LTGMs have scalable computational power to adapt to downstream tasks \cite{bommasani2021opportunities}, it can be expected to generate the new context of images with relative ease.
By doing so, visual artists who want to leverage LTGMs for their own purpose would use them to get inspired.
Furthermore, visual artists should be able to incorporate their identities into the generated images.
To this end, the interface could contain a place to upload the visual artists' portfolios and be customized with their own styles.

\subsection{More Controllability Using Multi-modal Input}

We found one of the fundamental limitations of LTGMs lies in the text prompting.
For example, some situations cannot be generated with just a few sentences.
To solve this, we suggest that it is important to give more controllability to users when generating images.
This also aligns with a previous finding that people want to lead the creation process \cite{oh2018lead}.
We found in the analysis that the visual artists think the interaction process between the artifact and themselves very seriously.
By having more interaction, we expect that users would have higher satisfaction when using LTGMs.

Moreover, as it is often impossible to express feelings through just texts, we need to provide multi-modal input such as hand gestures and voice, for example.
Currently, it is inefficient for both cases where the imagery in the visual artist's mind is vivid or abstract.
When the imagery is vivid, translating the image into words can be inefficient additional work.
In this case, the interface may have to accept a rough drawing, and convert it to a more complete and sophisticated one.
On the other hand, when the imagery is abstract, other input would be of help in describing what the visual artists want to express.
For example, a tone of voice can be considered in describing the visual artists' feelings.
Through this process, they would feel that they are taking the initiative in the creative work.




\subsection{Prompt Engineering for the Ease of Writing Text Input}

We found that not every visual artists are familiar with describing what they want to create through texts.
Sometimes they cannot come up with a proper expression for a certain image.
Moreover, LTGMs may not operate at the visual artists' intentions, the generated images not meeting their expectations.
Structuring the text prompt into forms decipherable to the machine can leave the visual artists exhausted.
Therefore, it is highly recommended to have a text prompt engineering tool that can recommend and revise the structure of text input.
Moreover, styles that can be generated by LTGMs can be listed on the side so that the users can choose an appropriate one, if they do not have a specific goal.




\section{Limitations and Future Work}


Our study contains several limitations.
First, we recruited most of the interview participants who are in their 20s and 30s.
We think that they are relatively young and open to new technologies like AI.
If we recruited more senior visual artists, e.g., having experiences in the domain for longer than 30 years, we might have extracted more practical shortcomings of LTGMs to be adopted in real-world scenarios of the working environment.
Second, we did not deal with the social impacts of LTGMs that has not yet been carefully examined.
For example, LTGMs are trained on a massive dataset which might include artworks that did not gain permission from the corresponding authors.
We think this is an important topic to be addressed in the future, similar to how Deepfake was addressed in the HCI community.
In addition, we found that those who were in the blind spot have concerns about the fast technological advancement.
For example, some of them expressed a serious fear of losing their jobs.
In the future, we plan to study the effect of the polarization in adopting new technologies and its solutions that the HCI community can provide for those who need support.





\section{Conclusion}

Our paper researched how visual artists would adopt LTGMs to support their creative works.
A systematic literature review of 72 papers on generative models was performed to comprehend the context in which they have been used in the HCI domain.
Based on the analysis, we conducted an interview study with 28 visual artists who cover 35 unique visual art domains to seek answers to our research questions.
The results showed that LTGMs can perform diverse roles including automating the creation process, helping the ideation process, and facilitating or arbitrating in communication.
We further discussed four design guidelines that future researchers can refer to in building intelligent user interfaces using LTGMs.



\section*{Acknowledgements}
This work was partly supported by Institute of Information \& communications Technology Planning \& Evaluation(IITP) grant funded by the Korea government(MSIT) (No.2019-0-00421, AI Graduate School Support Program(Sungkyunkwan University)), partly by Institute of Information \& communications Technology Planning \& Evaluation (IITP) grant funded by the Korea government(MSIT) (No.2019-0-00075, Artificial Intelligence Graduate School \seqsplit{Program(KAIST))}, and partly by the National Research Foundation of Korea(NRF) grant funded by the Korea government(MSIT) (No. \seqsplit{NRF-2019R1A2C2089062} and No. NRF-2019R1A2C1088900).

\bibliographystyle{ACM-Reference-Format}
\bibliography{9-ref}

\end{document}